\documentclass[showpacs,prd,twocolumn,floatfix,
nofootinbib,superscriptaddress]{revtex4}
\usepackage{graphicx}

\newcommand{\be}{\begin{equation}}
\newcommand{\ee}{\end{equation}}
\newcommand{\nl}{\nonumber \\}
\newcommand{\msb}{{\overline{\rm MS}}}
\newcommand{\psibar}{\overline{\Psi}}
\newcommand{\ord}{{\cal O}}
\newcommand{\sigmav}{\mbox{\boldmath$\sigma$}}
\newcommand{\Bv}{{\bf B}}

\begin{document}


\title{ Neutral $B$ Meson  
Mixing in Unquenched  Lattice QCD}

\author{Elvira G\'amiz}
\affiliation{Department of Physics,
University of Illinois, Urbana, IL 61801, USA}
\author{Christine T.\ H.\ Davies}
\affiliation{Department of Physics \& Astronomy,
University of Glasgow, Glasgow, G12 8QQ, UK}
\author{G.\ Peter Lepage}
\affiliation{Laboratory of Elementary Particle Physics,
Cornell University, Ithaca, NY 14853, USA}
\author{Junko Shigemitsu}
\affiliation{Department of Physics,
The Ohio State University, Columbus, OH 43210, USA}
\author{Matthew Wingate}
\affiliation{Department of Applied Mathematics and Theoretical Physics,
University of Cambridge, Cambridge CB3 0WA, UK.}

\collaboration{HPQCD Collaboration}
\noaffiliation


\begin{abstract}
We study $B_d$ and $B_s$ mixing in unquenched lattice QCD employing 
 the MILC collaboration gauge configurations that include $u$, 
$d$ and $s$ sea quarks based on the improved staggered quark (AsqTad) 
action and a highly improved gluon action. We implement the valence 
light quarks also with the AsqTad action and use the nonrelativistic NRQCD
 action for 
the valence $b$ quark. We calculate hadronic matrix elements 
necessary for extracting CKM matrix elements from experimental 
measurements of mass differences $\Delta  M_d$ and 
$\Delta M_s$.  
We find $\xi \equiv f_{B_s} \sqrt{\hat{B}_{B_s}} \, / \,
 f_{B_d} \sqrt{\hat{B}_{B_d}}
 = 1.258(33)$, 
$ f_{B_d} \sqrt{\hat{B}_{B_d}} = 216(15)$MeV and 
$ f_{B_s} \sqrt{\hat{B}_{B_s}} = 266(18)$MeV.  
We also update previous results for decay constants and obtain 
$ f_{B_d}  = 190(13)$MeV, 
$ f_{B_s}  = 231(15)$MeV and 
$ f_{B_s}  /  f_{B_d}  = 1.226(26)$.  
The new lattice results lead to
 updated values for the ratio of CKM matrix elements 
  $|V_{td}| \, / \, |V_{ts}|$
and for the Standard Model prediction for $Br(B_s \rightarrow \mu^+ \mu^-)$ 
with reduced errors. We determine $|V_{td}|/|V_{ts}| = 0.214(1)(5)$ and 
$Br(B_s \rightarrow \mu^+ \mu^-) = 3.19(19) \times 10^{-9}$.

\end{abstract}


\maketitle


\section{Introduction}
The mass differences $\Delta  M_s$ and $\Delta  M_d$ between 
the ``heavy'' and ``light'' mass eigenstates 
in the neutral $B$ meson system have now been measured very accurately 
leading to the possibility of a precise determination of 
the ratio of two Cabibbo-Kobayashi-Maskawa (CKM) matrix elements 
$|V_{td}| \, / \, |V_{ts}|$ \cite{pdg,cdf,d0}.
 This ratio is an important ingredient 
in fixing one of the sides of the ``Unitarity Triangle'',  and hence 
plays a crucial role in consistency checks of the Standard Model. 
  Reaching the goal of determining  $|V_{td}| \, / \, |V_{ts}|$ 
 from the experimental $\Delta \, M_q$'s,  however, requires theory input 
on hadronic matrix elements of certain four-fermion 
operators sandwiched between the $B_q$ and $\overline{B_q}$ 
states.  Information on such hadronic matrix elements 
can only be obtained if one has control over the strong interactions, QCD, 
in the nonperturbative domain.

\begin{table}
\caption{ Details of configurations employed. $N_{tsrc}$ is the number 
of time sources used per configuration and $N_{sm}$ the number of 
smearings on the heavy propagator including the unsmeared local 
case. All quark masses are given in the 
MILC collaboration normalization convention with
 $u_0 = \langle plaq. \rangle^{1/4}$.  Errors in $r_1/a$ are 
estimated to be at the 0.5\% level.
}
\begin{center}
\begin{tabular}{|c|c|c|c|c|c|c|}
\hline
Set &$r_1/a$   & $a u_0 m_{sea}$ &  $a u_0 m_{val}$ & $N_{conf}$& 
$N_{tsrc/sm}$  & size \\
\hline
\hline
C1  & 2.645 & 0.005/0.050 & 0.005   & 677 & 4/2 & $24^3 \times 64$ \\
&                   && 0.040   &&&  \\
C2  & 2.635 & 0.007/0.050 & 0.007   & 834 & 4/2 & $20^3 \times 64$ \\
&                   && 0.040   &&&  \\
C3  & 2.619 & 0.010/0.050 & 0.010   & 672 & 4/3 & $20^3 \times 64$ \\
&                   && 0.040   &&&  \\
C4  & 2.651 & 0.020/0.050 & 0.020   & 459 & 4/3 & $20^3 \times 64$ \\
&                   && 0.040   &&& \\
\hline
F1  & 3.701 & 0.0062/0.031 & 0.0062 & 547 & 4/2  & $28^3 \times 96$ \\
&                    && 0.031  &&&  \\
F2  & 3.721 & 0.0124/0.031 & 0.0124 & 534 & 4/2 & $28^3 \times 96$ \\
&                    && 0.031  &&&\\
\hline
\end{tabular}
\end{center}
\end{table}

\vspace{.1in}
\noindent
 In this article we use lattice QCD methods 
to calculate the hadronic matrix elements that appear in the 
Standard Model to describe neutral $B$ meson mixing.
Simulations are carried out on unquenched configurations created by
 the MILC collaboration \cite{milc1}.  These configurations include effects from 
vacuum polarization due to three light quark flavors, $up$, $down$ 
and $strange$ ($N_f = 2 + 1$ configurations, where $N_f$ equals 
the number of sea quark flavors).  
The $up$ and $down$ quark masses are set equal 
to each other.  
Table I lists the six different ensembles 
used, together with their characteristics such as 
 the number of configurations, 
sea quark masses in lattice units, the valence quark masses employed 
for each ensemble, 
number of time sources and 
the number of smearings for the $b$ quark propagators.  Information on the 
lattice spacing $a$ is presented in terms of the ratio $r_1/a$, 
where $r_1$ is obtained from the static potential and 
$r_1/a$ has been calculated by the MILC collaboration for each of 
their ensembles \cite{r1}.  The bare $b$ and $s$ quark masses have been fixed 
already in previous simulations of the $\Upsilon$ \cite{upsilon} 
and Kaon \cite{milc2} systems.  The MILC collaboration unquenched configurations 
have been created using the ``fourth root'' procedure to remove the 
four fold degeneracy of staggered fermions and some theoretical issues 
remain concerning the validity of this procedure.  Considerable progress has been 
made, however, in addressing this important issue \cite{fourthrt0} and several 
 recent reviews  \cite{fourthrt} summarize our current understanding of the 
situation.  In this work we assume that physical QCD is obtained in the continuum limit, 
as implied by existing evidence.

\vspace{.1in}
\noindent
In a previous article the HPQCD collaboration presented the first 
$N_f = 2 + 1$ unquenched results for $B_s$ meson mixing parameters, 
based on simulations on two out of the above 6 MILC ensembles 
(sets C3 and C4) \cite{bsmixing}. In the 
present work we broaden considerably the scope of our studies of 
$B$ mixing phenomena.  We generalize to include both $B_d$ 
and $B_s$ mixing and we use two sets of lattice spacings 
(the first 4 ensembles in Table I have $a \sim 0.12$fm and 
are called ``coarse'' whereas the last two have $ a \sim 0.09$fm 
and are refered to as ``fine'' lattices). Furthermore  we now employ 
smeared operators for the $B_q$ meson interpolating operators 
and even on those ensembles used previously we have doubled the 
statistics, by going from two to four time sources.  Unquenched lattice 
calculations by other groups exist in the literature. 
Several years ago the JLQCD collaboration published $N_f=2$ 
studies of $B_d$ and $B_s$ mixing \cite{jlqcd}
and the Fermilab/MILC collaboration has recently presented preliminary 
$N_f = 2+1$ results based on the same AsqTad light quarks as in the present 
article, however using a different action for the $b$ quarks \cite{tod,elvira}.

\vspace{.1in}
\noindent
In the next section we summarize the formulas needed for analysis 
of $B$ meson mixing phenomena.  We introduce the relevant four-fermion 
operators and describe how their matrix elements are 
parameterized and how they can be related to the CKM matrix elements 
$|V_{td}|$ and $|V_{ts}|$.  We then discuss the lattice four-fermion 
operators used in the simulations and how they can be matched 
onto the operators in continuum QCD.  In section III we describe our 
simulation data and the fitting procedures one must go through 
in order to extract the matrix elements of interest. Section IV focuses on
 chiral and continuum extrapolations and section V presents results for 
 $\xi \equiv f_{B_s} \sqrt{B_{B_s}} \, / \, f_{B_d} \sqrt{B_{B_d}}$,
 $f_{B_d} \sqrt{\hat{B}_{B_d}}$ and 
$f_{B_s} \sqrt{\hat{B}_{B_s}}$ together with discussions of systematic errors. 
This section, section V, summarizes the main 
results of the present work for quantities most directly associated with 
$B$ mixing analysis.  As part of our simulations, however, we have 
also accumulated more data on $B$ meson decay constants, 
$f_{B_d}$ and $f_{B_s}$.  
Hence in section VI we update the results for these decay constants 
published previously in \cite{fbprl,fbsprl}. 
   Section VII presents a summary of the current work 
   and a discussion of future directions
    in our program. 

\vspace{.1in}
\noindent
 We conclude this introductory section with a comment
    on notation. The decay constants $f_{B_q}$ with $q = d, \; s$
    are defined in eq.(2) below and are used together with appropriate
    bag parameters
    to parameterize four-fermion operator matrix elements.  $f_{B_d}$
    can of course be identified with $f_B$, the decay constant of the
    charged B mesons, and $f_B$ can be measured through the latter meson's
    leptonic decays. The $B_s$ meson, on the other hand, cannot decay
    leptonically via a single W boson  and hence $f_{B_s}$ by itself
    is not a directly measurable quantity in the Standard Model.
    Although $f_B$ is the more physical quantity we use the notation
    $f_{B_d}$ throughout this article in order to facilitate uniform
    treatment of $B_d$ and $B_s$ mixing.

\section{Relevant Four Fermion Operators and Matching}
Neutral $B$ meson mixing occurs at lowest order in the Standard Model 
through  box diagrams involving the exchange of two $W$ bosons.  
These box diagrams can be well approximated by an effective Hamiltonian 
expressed in terms of four-fermion operators. More specifically, for 
calculations of $\Delta M_q$ in QCD one is interested in the operator 
with  [V-A] x [V-A] structure,
\be 
\label{ol}
OL  \equiv  \left[\psibar_b^i(V-A) \Psi_q^i \right]\, 
\left [\psibar_b^j (V-A) \Psi_q^j \right]
\ee
where  $i$ and $j$ are color indices and are summed over. The symbol 
$q$ stands for either the $down$ or the $strange$ quark.  Working in the 
$\msb$ scheme, it is customary to parameterize 
the matrix element of $OL$ between a $B_q$ and a 
$\overline{B_q}$ state as,
\be
\label{defol}
\langle OL \rangle ^{\overline{MS}}(\mu)
\equiv
\langle \overline{B}_q | OL| B_q \rangle ^{\overline{MS}}(\mu)
\equiv \frac{8}{3}
f^2_{B_q} \, B_{B_q}(\mu)\,  M^2_{B_q}.
\ee
Here $f_{B_q}$ is the $B_q$ meson decay constant and 
$B_{B_q}$ its ``bag parameter''.  Factors such as 
 $\frac{8}{3}$ ensure that $B_{B_q} = 1$ in the ``vacuum 
saturation'' approximation.
Given the definitions in (\ref{defol}) the Standard Model prediction for 
the mass difference is \cite{buras},
\be
\label{deltamq}
\Delta M_q = \frac {G_F^2 M_W^2}{6 \pi^2} |V_{tq}V^*_{tb}|^2 \eta_2^B
S_0(x_t) M_{B_q} f^2_{B_q} \hat{B}_{B_q},
\ee
where $x_t = m_t^2/M_W^2$ depends on the $top$ quark and the $W$ boson 
masses $m_t$ and $M_W$,  $\eta_2^B$ is  a perturbative QCD correction
factor and  $S_0(x_t)$ the Inami-Lim function.  $\hat{B}_{B_q}$ 
is the renormalization group invariant bag parameter and at two-loop 
accuracy one has $\hat{B}_{B_q}/B_{B_q} = 1.539$ for the present case. 
From eq.(\ref{deltamq}) one sees that an 
experimental measurement of $\Delta M_q$ would yield directly the 
CKM matrix element combination $|V_{tq} V^*_{tb}|^2$  provided 
the quantity $f^2_{B_q} \hat{B}_{B_q}$ is available.
One also sees that the ratio $|V_{td}|/|V_{ts}|$ can be obtained from, 
\be
\frac{|V_{td}|}{|V_{ts}|} = \xi \sqrt{\frac{\Delta M_d}{\Delta M_s}
 \frac{M_{B_s}}{M_{B_d}}} \; , \qquad \xi \equiv \frac{f_{B_s}
\sqrt{B_{B_s}}}{f_{B_d} \sqrt{B_{B_d}}}.
\ee
The goal is to evaluate the hadronic matrix element in eq.(\ref{defol}) 
using lattice QCD methods.  Several steps are required in going from 
what is actually simulated on the lattice to the $\msb$ scheme quantities 
appearing in the continuum phenomenology formulas.
One important step is to relate four-fermion operators in continuum QCD 
to operators written in terms of the heavy and light quark fields appearing 
in the lattice actions that we employ.  Another crucial step will be to 
correct for the fact that simulations are carried out at nonzero lattice 
spacings and with light quark masses larger than the $up$ or $down$ quark masses 
in the real world.  In the remainder of  this section we address the first step, 
namely matching between the continuum QCD operator $OL$ and its counterpart 
in the effective lattice theory that we simulate.  The other step of 
chiral and continuum extrapolations will be discussed in section IV.

\vspace{.1in}
\noindent
Our simulations are carried out using the improved staggered (AsqTad) quark 
action for the light quarks \cite{stagg} and the nonrelativistic (NRQCD) action 
for the heavy quarks \cite{nrqcd}.
Matching through ${\cal O}(\alpha_s, \Lambda_{QCD}/M, 
\alpha_s/(aM))$ for the lattice 
action of this article was completed in reference \cite{fourfmatch}, 
 where $M$ is the heavy quark mass. 
We refer the reader to that paper 
for details and just summarize the most important formulas here. 
In effective theories such as NRQCD  one works separately with 
heavy quark fields  that create heavy quarks ($\psibar_Q$) and
with those that
annihilate heavy antiquarks ($\psibar_{\overline{Q}}$).  The  operator that 
contributes to $B_q - \overline{B_q}$ mixing at tree-level and 
that matches onto (\ref{ol}) at lowest order in $1/M$ has the form,
\begin{eqnarray}
OL^{eff}  & \equiv & \left[\psibar_Q^i(V-A) \Psi_q^i \right]\, 
\left [\psibar_{\overline{Q}}^j (V-A) \Psi_q^j \right]  \nl
&+&  \left[\psibar_{\overline{Q}}^i(V-A) \Psi_q^i \right]\, 
\left [\psibar_Q^j (V-A) \Psi_q^j \right]
\end{eqnarray}
As is well known, even at lowest order in 1/M 
there is a one-loop order mixing with another 
four-fermion operator, 
\begin{eqnarray}
OS^{eff}  &\equiv&  \left[\psibar_Q^i(S-P) \Psi_q^i \right]\, 
\left [\psibar_{\overline{Q}}^j (S-P) \Psi_q^j \right] \nl
&+&  \left[\psibar_{\overline{Q}}^i(S-P) \Psi_q^i \right]\, 
\left [\psibar_Q^j (S-P) \Psi_q^j \right]
\end{eqnarray}
This is true both in NRQCD and in HQET. 
If one introduces an effective theory field,
\be
\psibar^{eff}_b = \psibar_Q + \psibar_{\overline{Q}}
\ee
then $\psibar^{eff}_b$ and the QCD field $\psibar_b$ are related
by a Foldy-Wouthuysen-Tani (FWT) transformation.  In particular,
\be
\label{fwt}
\psibar_b = \psibar^{eff}_b \, \left [ I + \frac{1}{2M} \vec{\gamma} \cdot
\vec{\nabla}  \; + \; {\cal O}(1/M^2) \right ]
\ee
where the $\vec{\nabla}$ acts to the left. The FWT transformation determines the
 tree-level 1/M 
corrections to the four-fermion operators in the effective theory.  For 
$OL^{eff}$ they come in as,  
\begin{eqnarray}
\label{effj1}
OLj1 &=& \frac{1}{2M} \left [ \left (
 \vec{\nabla} \psibar_Q  \, \cdot \, \vec{\gamma} \, (V-A) \, \Psi_q \right )
 \left ( \psibar_{\overline{Q}} \, (V-A) \, \Psi_q \right ) \right . \nl
 &+& \left . \left (
 \psibar_Q \, (V-A) \, \Psi_q \right )
 \left ( \vec{\nabla}\psibar_{\overline{Q}}
\, \cdot \, \vec{\gamma} \, (V-A) \, \Psi_q \right ) \right ] \nl
 &+& \left [\psibar_{\overline{Q}} \rightleftharpoons \psibar_Q \right ] .
\end{eqnarray}
Taking  these corrections into account one can work  
through ${\cal O}(\alpha_s, \Lambda_{QCD}/M, \alpha_s/(aM))$ and finds 
the following  matching relation,
\begin{eqnarray}
\label{olmsbar}
&&\langle OL \rangle^{\overline{MS}}(\mu) = \nl
&& [ \, 1 + \alpha_s \, \rho_{11} \,]
 \, \langle OL^{eff} \rangle
  \, + \, \alpha_s \, \rho_{12} \, \langle OS^{eff} \rangle +  \nl
&  & \langle OLj1 \rangle
- \alpha_s \, \left [\, \zeta^{11} \,
\langle OL^{eff} \rangle \, + \,  \zeta^{12} \,
\langle OS^{eff} \rangle \, \right ] \nl
&& \; \; + \;{\cal O}(\alpha_s^2, \alpha_s \Lambda_{QCD} / M).
\end{eqnarray}
The matching coefficients $\rho_{11}$, $\rho_{12}$, $\zeta^{11}$ and 
$\zeta^{12}$ are listed (for $\mu = M_b$) in \cite{fourfmatch}.  As explained there,
the terms proportional to $\zeta^{ij}$ are needed to remove 
${\cal O}(\alpha_s/(aM))$ power law contributions in the matrix elements 
$\langle OLj1 \rangle$.

\section{Simulation Data and Fitting}
The starting point for a lattice simulation determination of 
$\langle \hat{O} \rangle$, with $\hat{O} = OL^{eff}$, $OS^{eff}$ or  
$OLj1$, is the calculation of the three-point correlator,
\begin{eqnarray}
\label{thrpnt}
  && C^{(4f)}_{\alpha \beta}(t_1,t_2) = \nl
 && \sum_{\vec{x}_1,\vec{x}_2} \langle 0 |
\Phi^{\alpha}_{\overline{B}_q}(\vec{x}_1,t_1) \; O^L(0)
\; \Phi^{\beta \dagger}_{B_q}(\vec{x}_2,-t_2) | 0 \rangle. \nl
\end{eqnarray}
One works with dimensionless operators $O^L \equiv a^6 \hat{O}$ which are 
kept fixed at the origin of the lattice. 
 $\Phi^{\alpha}_{B_q}$ is an interpolating operator for the $B_q$ meson 
 of smearing type ``$\alpha$'', and 
spatial sums over $\vec{x}_1$ and $\vec{x}_2$ ensure one is dealing with 
zero momentum $B_q$ and $\overline{B_q}$ incoming and outgoing states. 
The $B_q$ meson is created at time $-t_2$ and propagates to time slice $0$ where 
it mixes into a $\overline{B_q}$ meson. 
The $\overline{B_q}$ meson then propagates further in time 
until it is annihilated at time $t_1$.  We have accumulated data for 
$1 \leq t_1,t_2 \leq T_{max}$ with $T_{max} = 24$ on the 
coarse lattices and $T_{max} = 32$ on the fine lattices.  
Given the well known properties of staggered light quarks, 
for fixed $\alpha, \beta$ the 
three-point correlator must be fit to 
\begin{eqnarray}
\label{cf4f}
 && C^{(4f)}_{\alpha \beta}(t_1,t_2) =  \nl
 \;\;\;&& \sum_{j=0}^{N-1} \sum_{k=0}^{N-1}  A^{\alpha \beta}_{jk} \;
 e^{-E_j (t_1-1)}\; e^{-E_k (t_2-1)} \nl
&+& \sum_{j=0}^{\tilde{N}-1} \sum_{k=0}^{N-1}  B^{\alpha \beta}_{jk} \;
 (-1)^{t_1} \; e^{-\tilde{E}_j (t_1-1)}\; e^{-E_k (t_2-1)} \nl
&+& \sum_{j=0}^{N-1} \sum_{k=0}^{\tilde{N}-1}  C^{\alpha \beta}_{jk} \;
 (-1)^{t_2} \; e^{-E_j (t_1-1)}\; e^{-\tilde{E}_k (t_2-1)} \nl
&+& \sum_{j=0}^{\tilde{N}-1} \sum_{k=0}^{\tilde{N}-1}  D^{\alpha \beta}_{jk} \;
 (-1)^{t_1} (-1)^{t_2}\; e^{-\tilde{E}_j (t_1-1)}\;
 e^{-\tilde{E}_k (t_2-1)}.  \nl
\end{eqnarray}
This ansatz allows for $N$ non-oscillatory and $\tilde{N}$ oscillatory 
contributions to the correlator  (in practice we have  worked 
with $N = \tilde{N}$).  Not all the amplitudes $A^{\alpha \beta}_{jk}$ etc. 
are independent due to symmetries. 
Similarly two-point correlators are fit to,
\begin{eqnarray}
\label{twopnt}
 && C^{2pt}_{\alpha \beta}(t)  \equiv   \sum_{\vec{x}_1,\vec{x}_2} \langle 0 |
\Phi^{\alpha}_{B_q}(\vec{x}_1,t) 
\; \Phi^{\beta \dagger}_{B_q}(\vec{x}_2,0) | 0 \rangle \nl
 && =\sum_{j=0}^{N-1} b^\alpha_j b^\beta_j e^{-E_j (t-1)}
 + (-1)^t \sum_{k=0}^{\tilde{N}-1} 
\tilde{b}^\alpha_k \tilde{b}^\beta_k e^{-\tilde{E}_k (t-1)}. \nl
\end{eqnarray}
The relation between the amplitudes $A^{\alpha \beta}_{jk}$ or 
the $b^\alpha_j$  and the matrix elements of the 
previous section can be identified as follows.  
\be
A^{\alpha \beta}_{jk} = \frac{\langle 0 | \Phi^{\alpha}_{\overline{B_q}} 
|E_j\rangle \, \langle E_j | O^L| E_k \rangle \, \langle E_k | 
\Phi^{\beta \dagger}_{B_q} 
|0\rangle}{(2E_j a^3) (2 E_k a^3)}.
\ee
The energy eigenstates in the numerator are taken to have
 conventional  relativistic normalization and the factors in the denominator 
are needed to make up the difference between this continuum normalization 
and the one in the effective lattice theory.
For the ground state contribution $A^{\alpha \beta}_{00}$, and recalling that 
$O^L = a^6 \hat{O}$, one has,
\be
\label{a00}
A^{\alpha \beta}_{00} = \frac{\langle 0 | \Phi^{\alpha}_{\overline{B_q}} 
|\overline{B_q}\rangle \,\langle \overline{B_q} | \hat{O}|
 B_q \rangle \, \langle B_q | \Phi^{\beta \dagger}_{B_q} 
|0\rangle}{(2M_{B_q})^2}.
\ee
which includes the matrix element $\langle \overline{B_q} | \hat{O} | 
B_q \rangle$ that we are interested in.  Similarly for the 2pt-functions 
one has,
\be
\label{b0}
b^\alpha_0 b^\beta_0 = \frac{\langle 0 | \Phi^{\alpha}_{B_q} 
|B_q\rangle \,
\langle B_q | \Phi^{\beta \dagger}_{B_q} |0\rangle}{(2M_{B_q} a^3)}.
\ee
Using 
$\langle 0 | \Phi^{\alpha}_{\overline{B_q}} |\overline{B_q}\rangle  =
\langle 0 | \Phi^{\alpha}_{B_q} |B_q\rangle  $ one then has,
\be
 \langle \overline{B_q} | \hat{O}| B_q \rangle = \frac{2 M_{B_q}}{a^3} 
\frac{A^{\alpha \beta}_{00}}{b^\alpha_0 b^\beta_0} 
\ee
In order to assemble all the terms on the RHS of (\ref{olmsbar}) we have tried two 
approaches.  In the first approach we did separate fits for each of the operators 
$\hat{O} = OL^{eff}$, $OS^{eff}$ and $OLj1$ and inserted their ground state
 matrix elements into (\ref{olmsbar}). In the second approach we went through 
the analysis in the opposite order. Namely we first obtained the renormalized 
four-fermion operator at the three-point function level by forming the 
appropriate linear combinations of the $C^{(4f)}$'s, and then carried out 
fits to extract $A_{00}$ for the full renormalized three-point function. 
  Consistent results were obtained 
from the two methods.  For our final analysis we adopted the second 
approach which we found to be more convenient in practice.

\begin{figure}
\includegraphics*[width=8.0cm,height=9.0cm,angle=270]{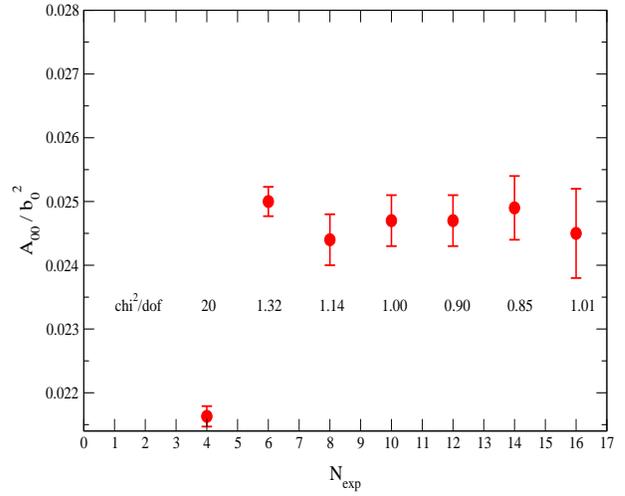}
\caption{
Fit results for $\frac{A_{00}}{b_0^2}$ versus the number of 
exponentials $N_{exp} = N + \tilde{N}$ for one of the coarse ensembles,
Set C2 with $au_0 m_{val} = 0.04$.
 }
\end{figure}
\begin{figure}
\includegraphics*[width=6.3cm,height=8.0cm,angle=270]{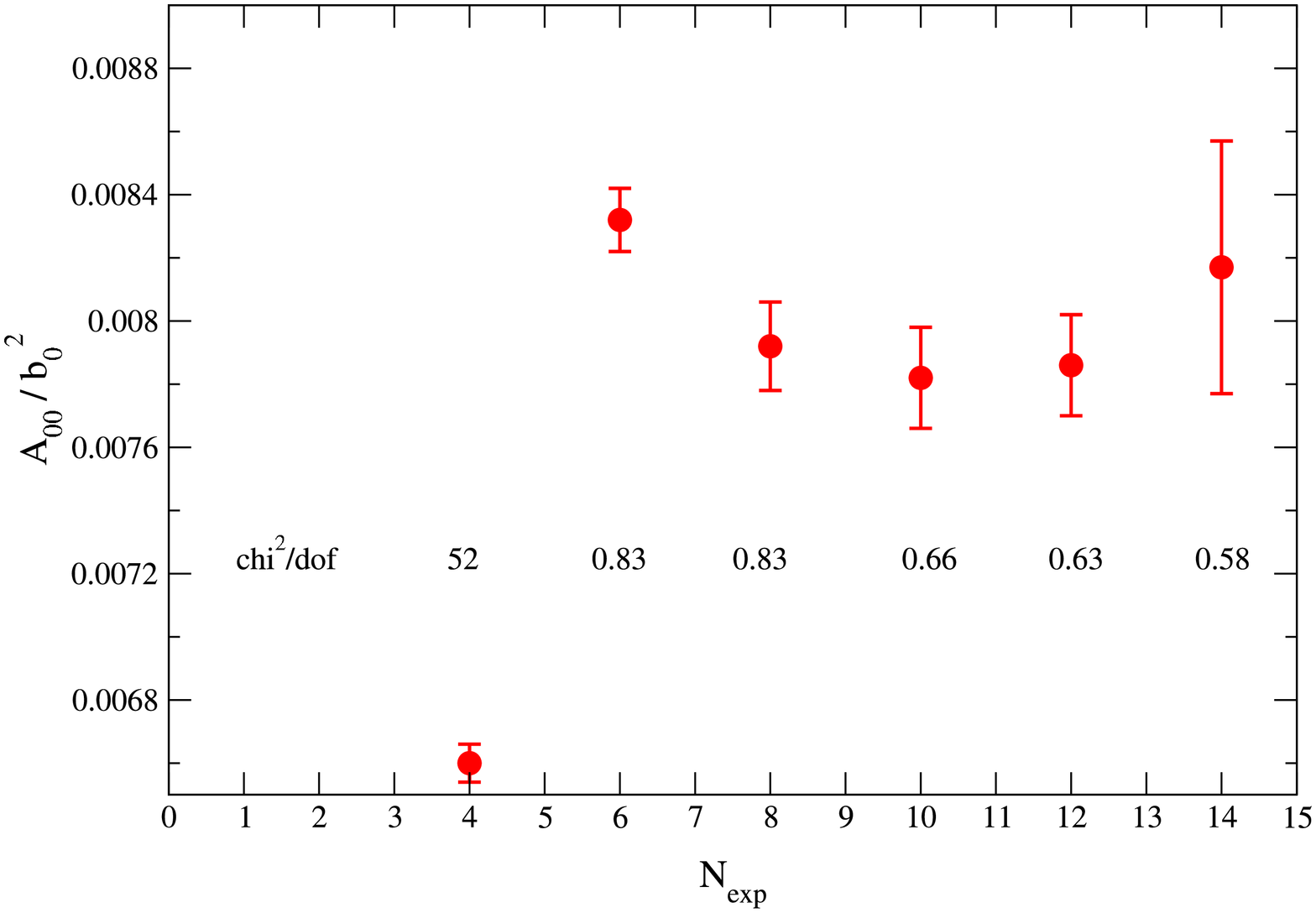}
\caption{
Same as Fig.1 for a fine ensemble, Set F1 with $au_0 m_{val} = 0.031$.
 }
\end{figure}

\noindent
Our smearings consist of Gaussian smearings of the heavy quark 
propagator at both source and sink.  
In addition to point sources and sinks we use Gaussians with widths, in units 
of the lattice spacing, of 2.0 and 6.0 for sets C1 and C2 and  one Gaussian 
each of width 6.0 for C3 and C4 and of width 8.0 for sets F1 and F2.  
To extract  $\frac{A_{00}}{b_0^2}$ for our renormalized 
three-point function we  carry out 
simultaneous fits to an
 $ N_{sm} \times N_{sm}$  matrix of two-point correlators (eq.(\ref{twopnt}) 
with $\alpha, \beta = 1,... N_{sm}$) ($\alpha=1$ corresponds to local, 
$\alpha=2$ to first Gaussian etc.)
and to the renormalized 
three-point functions  with $\alpha = \beta$.  
 Bayesian fitting \cite{bayes} 
methods are employed to enable these complicated fits with large numbers of 
exponentials, i.e. of fit parameters.
We fit to all data points 
within $t_{min} \leq t \, , \, t_1 \leq t_{max}$ and $t_{min} \leq t_2 \leq 
t^\prime_{max}$  for $t_{min} = 2 \sim 3$, 
$t_{max} = 20 \sim 24$ and $t_{max}^\prime = 13 \sim 15$.
  We have used $N_{exp} \equiv N + \tilde{N}$
 ranging 
between 4 to 16 and looked for consistency in fit results   
as the number of exponentials was increased.  An example of fit results on
one of the coarse lattices is shown in Fig.1.  One sees that good and 
consistent results are obtained for $8 \leq N_{exp} < 16$.  
When $N_{exp}$ becomes very large (in the case of Fig.1 $\geq 16$), 
errors tend to increase again indicating that our fit ansatz has become too 
complicated for the minimization routines to handle, given the amount
of statistics that we have.
Fig.2 shows an example for one of the fine lattices.  Here we find good 
 results for $8 \leq N_{exp} < 14$.  In general we have relied 
on our $N_{exp}=8,10$ and $12$ fits for all our ensembles.

\vspace{.1in}
\noindent
 We summarize fit results in Table II.  The dimensionful quantities 
$f_{B_q} \sqrt{M_{B_q} \hat{B}_{B_q}}$ are given in units of $r_1^{-3/2}$.  
Errors include both statistical plus fitting errors and errors coming from 
uncertaintiy in $r_1/a$ which we take to be $\sim 0.5$\%.
  Note that we have also gone to the 
renormalization group invariant bag parameter $\hat{B}_{B_q}$.

\begin{table}[tb!]
\caption{ Fit results for $f_{B_q} \sqrt{M_{B_q} \hat{B}_{B_q}}$ in 
units of $r_1^{-3/2}$ and for the dimensionless ratio 
$\xi \sqrt{\frac{M_{B_s}}{M_{B_d}}} = 
\frac{f_{B_s}\sqrt{B_{B_s} M_{B_s}}}{f_{B_d} \sqrt{B_{B_d} M_{B_d}}}$. 
Errors in the last column are statistical + fitting errors.  Those in the 
second and third columns include additional errors coming from the 
0.5\% uncertainty in $r_1/a$.
}
\begin{center}
\begin{tabular}{|c|c|c|c|}
\hline
Set  & $r_1^{3/2} f_{B_s} \sqrt{M_{B_s} \hat{B}_{B_s}}$ &
 $r_1^{3/2} f_{B_d} \sqrt{M_{B_d} \hat{B}_{B_d}}$ & $\xi \sqrt{\frac{M_{B_s}}
{M_{B_d}}}$  \\
\hline
\hline
C1 & 1.430(21)  &  1.193(27)   & 1.199(29) \\
C2 & 1.442(16)  &  1.248(35)   & 1.155(33) \\
C3 & 1.382(21)  &  1.179(21)   & 1.172(23) \\
C4 & 1.413(18)  &  1.263(22)   & 1.119(22) \\
\hline
F1 & 1.353(17)  &  1.138(28)   & 1.189(26)  \\
F2 & 1.334(20)  &  1.193(27)   & 1.118(24)  \\
\hline
\end{tabular}
\end{center}
\end{table}

\section{Chiral and Continuum Extrapolations }

The lattice data presented in Table II are for simulations with 
$up$ and $down$ quark masses $m_u = m_d$ larger than 
in the real world and need to be extrapolated to the physical point. Reaching 
this physical point also involves taking the lattice spacing $a \rightarrow 0$ 
limit.  We use staggered chiral perturbation theory (SChPT) \cite{schpt1, 
schpt2, schpt3} 
 augmented by further 
general discretization correction terms to carry out the simultaneous 
chiral and continuum extrapolations.  Continuum heavy meson chiral perturbation
for $B$ and $D$ mixing was developed in \cite{detlin,kamenik1} including for the 
partially quenched case. These formulas were generalized recently to 
next-to-leading order SChPT  by Bernard, Laiho and Van de Water \cite{schpt4} 
 and  generously made available to us prior to publication.
We use the following fit ansatz,
\begin{eqnarray}
\label{schpt1}
 && r_1^{3/2} \, f_{B_q}\sqrt{M_{B_q} \hat{B}_{B_q}} =  \nl 
  && c_1 \, [1 +  \frac{1}{2} \,\Delta f_q   + 
c_2 \, (2 m_f + m_s) \, r_1 +  c_3 \, m_q \, r_1 ] \times  \nl
 &&[ 1 + c_4 \,  
 \alpha_s ( a/r_1 )^2  + c_5 \,  ( a/r_1 )^4 ] .
\end{eqnarray}
$\Delta f_q$ stands for  the chiral log contributions and includes
 the staggered light quark action specific taste breaking terms. The factor 
of $1/2$ comes about since $\Delta f_q$ was calculated for the square, 
namely for $f^2_{B_q} M_{B_q} B_{B_q}$. We use the 
notation $m_f$ and $m_s$ for the sea $up/down$ and $strange$ quark masses 
respectively, and $m_q$ (or $m_{qs}$) for the valence quark masses.
The second bracket parametrizes further discretization corrections that 
are expected to come in at ${\cal O}(\alpha_s \, a^2)$ and ${\cal O}(a^4)$.  
We have also tried adding more analytic terms with higher powers of 
quark masses. 

\noindent
 $\Delta f_q$ includes the coupling $g_{B^*B \pi}$ which has 
not been measured experimentally.  However, based on Heavy Quark Effective 
Theory (HQET) arguments, $g_{B^*B \pi}$ is believed to be close to an 
analogous coupling $g_{D^*D \pi}$ in the $D$ meson system for which some 
experimental information is available. The latter coupling is estimated to be 
between $0.3 \leq g_{D^*D \pi} \leq 0.6$ \cite{gddpi}. 
 As we discuss below, we have carried out 
two types of fits, one  
where we did a whole sequence of fits with $g_{B^*B\pi}$ varying between 
$0 \leq g_{B^*B \pi} \leq 0.6$ but where this coupling was kept fixed during 
each individual fit.  In the second type of fit we let the coupling float and 
be one of the fit parameters.  Both types of fits favored smaller values 
with $g_{B^*B\pi} \approx 0.1$, however as long as $g_{B^*B\pi} < 0.5$ 
fit results were quite insensitive to its exact value.

\noindent
For the ratio $\xi \sqrt{M_{B_s}/M_{B_d}}$ we use,
\begin{eqnarray}
\label{schpt2}
&& \xi  \, \sqrt{\frac{M_{B_s}}{M_{B_d}} }=    \nl
&&  [1 +  \frac{1}{2} \,(\Delta_0 f_{qs} - \Delta_0 f_q)  + 
  \frac{b_1^2}{2} \,(\Delta_1 f_{qs} - \Delta_1 f_q)  +  \nl
 &&  b_2 \, (m_{qs} - m_q) \, r_1 +  b_3 \,(m_{qs} -  m_q)^2 \, r_1^2] \times  \nl
 &&[ 1 + ( b_4 \,  
 \alpha_s ( a/r_1 )^2  + b_5 \,  ( a/r_1 )^4) (m_{qs} - m_q)\, r_1 ] . 
\end{eqnarray}
Here we have split up,
\be
\Delta f_q = \Delta_0 f_q + g^2_{B^*B\pi} \, \Delta_1 f_q
\ee
and then let $g_{B^*B\pi} \rightarrow b_1$ become one of the fit parameters. 
In Fig.3 we show a simultaneous fit to the six entries in the last 
column of Table II. The green and blue curves are the curves from this fit 
appropriate to the  coarse and 
fine lattice data points respectively and the red curve is the ``continuum'' curve 
 obtained by retaining the fitted values for $b_1$, $b_2$ and $b_3$ and turning off the 
$b_4$ and $b_5$ correction terms plus the taste breaking contributions inside 
 $\Delta f_q$ and $\Delta f_{qs}$. One sees that within our statistical and fitting 
errors of $\sim 2$\%, there is consistency between the three curves.  In other words, 
we see almost no statistically significant 
lattice spacing dependence in this ratio. At the physical point 
the difference between the green and blue curves is 1.8\%, which reduces to
1.3\% if the green curve is adjusted and corrected for having a 
sea $strange$ quark mass on the coarse lattices that is about 20\% too large.  
One might be surprised that the magenta curve lies below the blue curve.  This 
comes about because  the various discretization effects inside 
$(\Delta f_{qs} - \Delta f_q)$ 
and in the $b_4$ \& $b_5$ terms can have different signs and come in with 
 different relative weights between the coarse and fine 
lattices.  All these effects come in at the $\sim0.5$\% or less level, 
and are hence too 
small to allow us  to disentangle one from 
the other in a meaningful way. 
The fit shown in Fig.3 has $\chi^2_{aug}/dof = 0.54$ \cite{bayes2} 
 and gives $g_{B^*B\pi} = 0.14(47)$.

\begin{figure}
\includegraphics*[width=8.5cm,height=7.0cm]{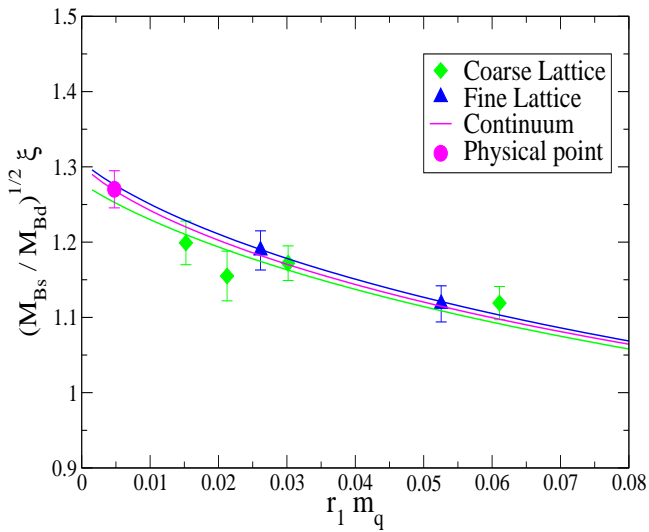}
\caption{
Chiral and continuum extrapolation of the ratio $\xi 
\sqrt{\frac{M_{B_s}}{M_{B_d}}}$.  Errors shown are statistical plus fitting 
errors. 
 The physical point is at $r_1  m_{qs}/27.4$, where $m_{qs}$ is the 
valence strange quark mass.
 }
\end{figure}

\vspace{.1in}
\noindent
 Fig.4 shows chiral \& continuum extrapolation curves for $r_1^{3/2} f_{B_d} 
\sqrt{\hat{B}_{B_d} M_{B_d}}$ using the fit ansatz of eq.(\ref{schpt1}).  
  Again the green and blue full curves 
are the fit curves
 for the coarse and fine lattice data respectively, and the dotted lines
 show the error bands around these central curves.  Turning off the $c_4$ and $c_5$ 
contributions and the taste breaking terms inside $\Delta f_q$ leads to 
the red curve which can be followed down to the physical point.  
In contrast to the situation for the ratio $\xi$, here, with $f_{B_d} \sqrt{
\hat{B}_{B_d} M_{B_d}}$, one finds a noticeable shift between the coarse and 
fine lattice points. The difference between the green and blue curves is 
a 5.5\% effect.  Going from the fine (blue) curve to the red continuum 
extrapolated curve is a 4\% shift, which is also the size of the
chiral \& continuum extrapolation  error at the physical point.  The fit in 
Fig.4 has $\chi^2_{aug}/dof = 0.99$.

\begin{figure}
\includegraphics*[width=8.5cm,height=7.0cm]{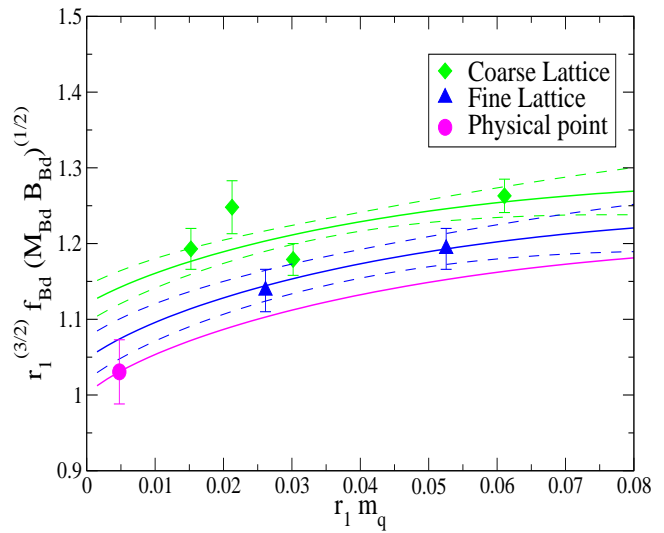}
\caption{
Chiral and continuum extrapolation of $r_1^{3/2} f_{B_d} \sqrt{\hat{B}_{B_d} M_{B_d}}$. 
Errors on the data points 
are statistical plus fitting errors combined with uncertainty in $r_1/a$.
The dashed curves correspond to the error bands about the central 
green and blue full lines.
 The physical point is at $r_1  m_{qs}/27.4$, where $m_{qs}$ is the 
valence strange quark mass.
 }
\end{figure}

\noindent
Finally, in Fig.5 we show results for $ r_1^{3/2} f_{B_s} \sqrt{\hat{B}_{B_s} M_{B_s}}$, 
where $\chi^2_{aug}/dof = 0.96$ for the simultaneous fit to all the data points.  Here 
 the difference between the green and blue curves is a 6\% effect and 
between the blue and red curve a 5.7\% effect.  These shifts are slightly larger than but 
similar to those for $B_d$ in Fig.4.
In both cases the discretization effects we are seeing in $r_1^{3/2} f_{B_q} 
\sqrt{\hat{B}_{B_q} M_{B_q}}$ are larger than the naive expectation of 
a leading correction of ${\cal O}(a^2 \alpha_s)$ which would be $\sim 4$\% or 
$\sim 2$\% on the coarse or fine lattices respectively.  It was hence very 
 important to have 
 simulations results at more than one lattice spacing and carry out an 
explicit continuum extrapolation.  Fortunately, for the important ratio
$\xi$ these discretization corrections  cancel out to a large extent, as 
expected and as we have already verified in Fig.3.

\begin{figure}
\includegraphics*[width=8.5cm,height=7.0cm]{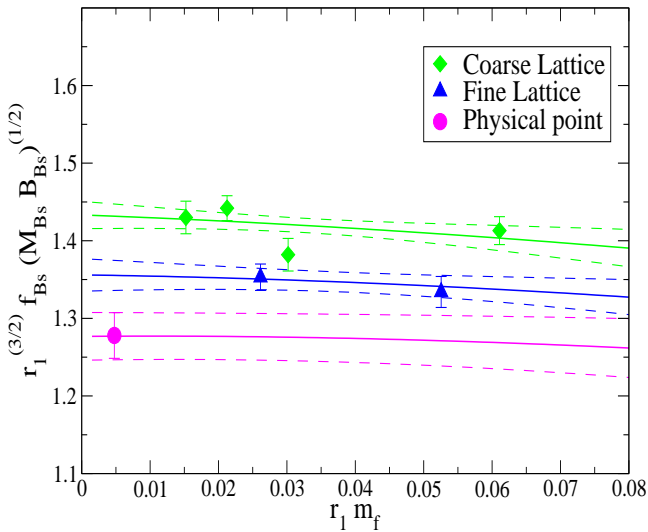}
\caption{
Same as Fig.4 for $r_1^{3/2} f_{B_s} \sqrt{\hat{B}_{B_s} M_{B_s}}$ versus $r_1 m_f$. 
 }
\end{figure}

\section{ Main Results and Error Budget}
Table III gives our error budget for the main uncertainties
in the three quantities, 
$f_{B_s} \sqrt{\hat{B}_{B_s}}$, 
$f_{B_d} \sqrt{\hat{B}_{B_d}}$ and $\xi$.

\begin{table}
\caption{ Errors in \% for 
$f_{B_s}\sqrt{\hat{B}_{B_s}}$, $f_{B_d}\sqrt{\hat{B}_{B_d}}$ and $\xi$.
}
\begin{center}
\begin{tabular}{|c|c|c|c|}
\hline
source of error & $f_{B_s} \sqrt{\hat{B}_{B_s}}$ &
 $f_{B_d} \sqrt{\hat{B}_{B_d}}$ & $\;\;\xi \;\;$ \\
\hline
\hline
stat. + chiral extrap.  & 2.3  &  4.1 &  2.0 \\
residual  $a^2$ extrap. &3.0  & 2.0 &  0.3\\
uncertainty &&&  \\
\hline
 $r_1^{3/2}$ uncertainty  & 2.3  &  2.3  &  --- \\
 $g_{B^*B\pi}$ uncertainty  & 1.0  &  1.0  &  1.0 \\
$m_s$ and $m_b$ tuning &  1.5  & 1.0  & 1.0  \\
operator matching  & 4.0 & 4.0& 0.7 \\
relativistic corr.  & 2.5 &  2.5 & 0.4\\
\hline
Total  & 6.7 & 7.1 & 2.6 \\
\hline
\end{tabular}
\end{center}
\end{table}

\begin{figure}
\includegraphics*[width=7.3cm,height=8.0cm,angle=270]{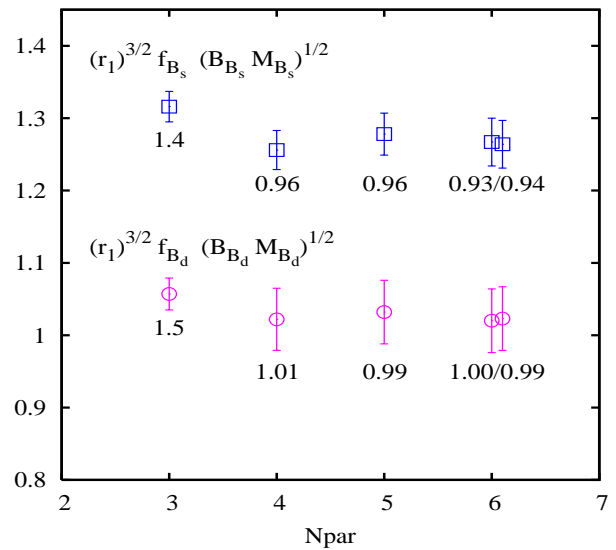}
\caption{
 $r_1^{3/2} f_{B_q} \sqrt{\hat{B}_{B_q} M_{B_q}}$ 
at the physical point versus $Npar$, the number of fit parameters.  
The numbers below the data points give $\chi^2_{aug}/dof$. For $Npar=6$ 
we give results for two different types of term added, 
a term proportional to the quark mass squared (right data point) or 
a further discretization correction $\propto (a/r_1)^3 \alpha_s$ 
(left data point).
 }
\end{figure}

\noindent
We explain each entry in Table III in turn.

\begin{itemize}
\item
{\it statistics and chiral extrapolations:} These are 
the errors shown on the ``physical points'' in Figs.3, 4 and 5 and 
are outputs from  our chiral \& continuum extrapolation fits.

\item
{\it residual $a^2$ extrapolation error:}  It is necessary to list this error 
separately since the degree to which the red curves in the above figures actually 
correspond to the true continuum limit depends on how well one has modelled 
discretization errors in our simulations. In other words one needs to 
assess the error in the fit ansatz for the continuum extrapolation (we assume 
the chiral extrapolation is handled sufficiently accurately by Staggered 
ChPT) and this turns out to be a nontrivial task. \\
On the one hand the data 
appears to be consistent with the fit ans\"atze of eqs.(\ref{schpt1}) and
(\ref{schpt2}).  We have tried adding further terms and found that 
fit results shifted by an amount less than, and in most cases much less than,
 the ``statistical + chiral extrapolation'' errors. Fig.6 shows results 
for the chirally and continuum extrapolated $r_1^{3/2} f_{B_q} 
\sqrt{\hat{B}_{B_q} M_{B_q}}$ versus the number of parameters $Npar$ 
in the fit ansatz.  
For $Npar=3$ one has discretization corrections only through the $\Delta f_q$ 
$S\chi PT$ term and one sees that a good fit to the data points at the two 
lattice spacings cannot be obtained.  However, once 
 $Npar > 3$, good fits are achieved and results and their errors
 are stable with respect to 
changes in $Npar$.\\
On the other hand we know that due to our 
use of the NRQCD action to describe the heavy $b$ quark, coefficients such as 
$c_4$ are in general complicated functions of $(aM)$ 
(although $c_4$ does  
include a constant piece coming from the light quarks).  \\
We have approached  this complicated situation in the following way.
We interpret the red ``continuum'' curves in Figs.3, 4 and 5 as the curves one 
would get after taking care of all discretization errors coming from the 
light quark and the glue sectors.  Then under ``residual $a^2$ extrapolation 
uncertainty'' one would include errors coming from the heavy quark action.  
The leading such error in our calculations is of 
${\cal O}(a^2 \alpha_s)$  multiplied 
by some function of $(aM)$ which we initially take to be of ${\cal O}(1)$ 
 leading to an additional uncertainty of $\sim2$\% in $f_{B_q} 
\sqrt{\hat{B}_{B_q}}$.  This would correspond to 
 a standard power counting estimate of discretization errors where one 
takes coefficients of order one in higher order corrections.  We have opted 
to be slightly more conservative in our power counting assessment and apply 
a factor of 1.5 rather than 1.0 for those cases where the ``physical'' 
(the magenta) points deviate by more than one $\sigma$ from the fine 
lattice (blue) curve. By ``$\sigma$'' we mean here the 
``statistical + chiral extrapolation'' errors.
 For $\xi$ 
we have multiplied the error for $f_{B_d} \sqrt{\hat{B}_{B_d}}$ 
by a factor of $\frac{(m_s - m_{u/d})}{\Lambda_{QCD}} \sim 1/6$.\\
Since the second row in Table III gives an assessment of the uncertainty 
in our $a^2$ extrapolation rather than an estimate of the full discretization error, 
we believe the procedure outlined here to fix it is a 
reasonably conservative one.

\item
{\it  $r_1^{3/2}$ uncertainty:} follows from the 1.5\% error in current 
determinations of the physical value for $r_1$.

\item
{\it uncertainty in $g_{B^*B\pi}$:} we  carried out fixed coupling chiral fits 
for the range $0.0 < g_{B^*B\pi} < 0.6$ and looked
 at the spread in the results at the 
physical point.  For couplings larger than 0.6, $\chi^2/dof$ starts to 
deteriorate.

\item
{\it tuning of strange and bottom quark masses:} The largest mistuning, which 
occurs in the sea strange quark mass $m_s$ on the coarse lattices, 
has been corrected for 
when calculating fit results at the physical point and residual effects have been
 estimated by varying this adjusted value for $m_s$. Errors due to uncertainty 
in the valence $strange$ quark mass have been assessed by comparing  $f_{B_q} 
\sqrt{\hat{B}_{B_q}}$ as one goes from valence quark mass $m_{qs}$ down to 
$m_q$ and 
  errors coming from mistuning 
of $m_b$ have been estimated from the $1/M$ dependence
 of decay constants studied in \cite{fbsprl}.

\item 
{\it operator matching} and {\it relativistic corrections}: These two 
sources of error are intimately intertwined and again how to separate
 the two is not clear cut.  As indicated in eq.(\ref{olmsbar}), 
our matching for $f^2_{B_q} \hat{B}_{B_q}$ has been carried out 
up to correction of ${\cal O}(\alpha_s^2)$ and ${\cal O}(\alpha_s 
\Lambda_{QCD}/M)$.  In Table III we have listed the first correction 
under ``operator matching'' and the latter correction under 
``relativistic corrections''.  And again the errors for $\xi$ are 
reduced by a factor of 1/6 relative to those for the two 
non ratio quantities. 

\end{itemize}

\noindent
Using central values coming from the physical (red) points in the
 figures and the 
errors summarized in Table III,  we can now present our main results.
\be
\label{xires}
\xi \equiv \frac{f_{B_s} \sqrt{B_{B_s}}}{f_{B_d} \sqrt{B_{B_d}}} 
= 1.258(25)(21),
\ee
and using $r_1 = 0.321(5)fm$ \cite{upsilon},
\be
\label{fbsres}
f_{B_s} \sqrt{\hat{B}_{B_s}} = 266(6)(17) \;\left (\frac{0.321}{r_1[fm]} \right)^{3/2} 
{\rm MeV},
\ee
\be
\label{fbdres}
f_{B_d} \sqrt{\hat{B}_{B_d}} = 216(9)(12) \;\left (\frac{0.321}{r_1[fm]} \right)^{3/2} 
{\rm MeV },
\ee
where the first error comes from statistics + chiral extrapolation and 
the second is the sum of all other systematic errors added in quadrature.  
From the individual $f_{B_q} \sqrt{\hat{B}_{B_q}}$, q=s or d, 
one obtains a ratio of 1.231(58)(21) which is consistent with (\ref{xires}) 
however with larger errors. The result for $f_{B_s} \sqrt{\hat{B}_{B_s}}$ in 
eq.(\ref{fbsres}) is consistent with but more accurate than our
previously published  value of $281(21)$MeV \cite{bsmixing}.

\section{Updates on  $f_{B_d}$, $f_{B_s}$ and $f_{B_s} \, / \, f_{B_d}$ and 
Estimates of Bag Parameters}

The numerical simulations of two-point and three-point functions,
 such as in  eqns.(\ref{twopnt}) and 
(\ref{thrpnt}), that enabled us to extract the B-mixing parameters of 
the previous section also provide information necessary to determine 
$B_d$ and $B_s$ meson decay constants $f_{B_d}$ and $f_{B_s}$. Decay constants 
are defined through the matrix element of the heavy-light axial vector 
current between the $B_q$ meson state and the hadronic 
vacuum. Using the temporal component $A_0$ and working in the heavy meson 
rest frame one has,
\be
\langle 0| A_0 |B_q \rangle \equiv M_{B_q} f_{B_q} .
\ee
Just as with the four-fermion operators of section II, matching is required 
between the heavy-light current in continuum QCD and currents made out of 
quark fields of the effective lattice theory.  This matching has been carried out 
at the one-loop order for NRQCD/AsqTad currents in \cite{pert1} based on formalism 
developed in \cite{pert2}.
\begin{eqnarray}
\label{a0}
\langle \, A_0 \, \rangle^\msb  &=& ( 1 + \alpha_s \,
\tilde{\rho}_0)\,\langle J^{(0)}_0 \rangle + \nonumber \\
 & & (1 + \alpha_s   \,  \rho_1) \, \langle
J^{(1),sub}_0 \rangle  
  \; + \; \alpha_s  \,
 \rho_2 \, \langle J^{(2),sub}_0 \rangle \nonumber \\
& + &  \quad \ord(\alpha_s^2, \Lambda_{QCD}^2/M^2)  \, ,
\end{eqnarray}
The heavy-light currents $J_0^{(i)}$ in the effective theory are defined as,
\begin{eqnarray}
\label{j0}
 J^{(0)}_{0} & = &  \psibar_q\,\Gamma_0\, \Psi_Q,  \\
\label{j1}
 J^{(1)}_{0} & = & \frac{-1}{2 \,(aM)} \psibar_q
    \,\Gamma_0\,\mbox{\boldmath$\gamma\!\cdot\!\nabla$} \, \Psi_Q, \\
\label{j2}
 J^{(2)}_{0} & = & \frac{-1}{2 \,(aM)}  \psibar_q
    \,\mbox{\boldmath$\gamma\!\cdot\!\overleftarrow{\nabla}$}
    \,\gamma_0\ \Gamma_0\, \Psi_Q.
\end{eqnarray}
with $\Gamma_0 \equiv \gamma_5 \gamma_0$ and 
\be
\label{jsub}
 J_0^{(i),sub} \equiv J_0^{(i)} - \alpha_s \,   \zeta_{i0}  \,
J_0^{(0)}
\ee
The matching coefficients $\rho_i$ and $\zeta_{i0}$ are given in \cite{pert1}. 
Note that the matching for the heavy-light current includes contributions 
at $\ord (\alpha_s \frac{\Lambda_{QCD}}{M})$ and hence is more accurate than the 
matching in (\ref{olmsbar}) for the four-fermion operator.

\begin{table}[b]
\caption{ Fit results for $\Phi_q = f_{B_q} \sqrt{M_{B_q}}$ in 
units of $r_1^{-3/2}$ and for the  ratio $\Phi_s / \Phi_d$. 
Errors are as described in Table II. 
}
\begin{center}
\begin{tabular}{|c|c|c|c|}
\hline
Set  & $ \quad r_1^{3/2} \Phi_s \quad$ &
 $\quad r_1^{3/2} \Phi_d \quad$ & $\quad \Phi_s / \Phi_d \quad$  \\
\hline
\hline
C1 & 1.261(12)  &  1.085(14)   & 1.162(14) \\
C2 & 1.246(11)  &  1.073(14)   & 1.162(12) \\
C3 & 1.236(12)  &  1.071(14)   & 1.155(14) \\
C4 & 1.248(16)  &  1.128(17)   & 1.107(20) \\
\hline
F1 & 1.175(13)  &  0.990(22)   & 1.188(20)  \\
F2 & 1.180(13)  &  1.047(16)   & 1.120(11)  \\
\hline
\end{tabular}
\end{center}
\end{table}

\noindent
We have evaluated the two-point functions,
\be
\label{jtwopnt}
  C^{2pt}_{j \beta}(t)  =   \sum_{\vec{x}_1,\vec{x}_2} \langle 0 |
J_0^{(j)}(\vec{x}_1,t) 
\; \Phi^{\beta \dagger}_{B_q}(\vec{x}_2,0) | 0 \rangle 
\ee
for $j = 0,1,2$.  We then calculate the renormalized current 
matrix element by forming the appropriate linear combination as dictated 
by the RHS of (\ref{a0}). This is done for both $B_d$ and $B_s$. 
The next step is to fit the renormalized two-point correlator using the ansatz of 
eq.(\ref{twopnt}), extract the relevant ground state matrix element 
and thereby obtain $\Phi_q \equiv f_{B_q} \sqrt{M_{B_q}}$.  
We do simultaneous fits to $B_d$ and $B_s$ 
correlators, so that $\Phi_d$, $\Phi_s$ and the ratio $\Phi_s/\Phi_d$ 
are determined within the same fit.  Fit results are summarized in 
Table IV. For $r_1^{3/2} \Phi_q$ errors include the uncertainty 
in $r_1/a$ in addition to statistical and fitting errors.

\noindent
The rest of the analysis for $\Phi_q$ and $\Phi_s/\Phi_d$ is very 
similar to what was done for the four-fermion operator matrix 
elements in section IV.  Chiral and continuum extrapolations are 
carried out using a fit ansatz of the form (\ref{schpt1}) for 
$r_1^{3/2} \Phi_q$ and (\ref{schpt2}) for $\Phi_s / \Phi_d$.  The only 
difference is that here $\Delta f_q$ will involve the chiral logarithms 
appropriate for decay constants rather than for four-fermion operators.  
Such contributions were calculated by Aubin \& Bernard using 
Staggered ChPT in reference \cite{schpt3}.  Fig.7, 8 and 9
 show chiral and continuum 
extrapolations for $\Phi_s/\Phi_d$, $\Phi_d$ and $\Phi_s$
with $\chi^2_{aug}/dof$ = 1.00, 1.06 and 0.53 respectively.

\begin{figure}
\includegraphics*[width=8.5cm,height=7.0cm]{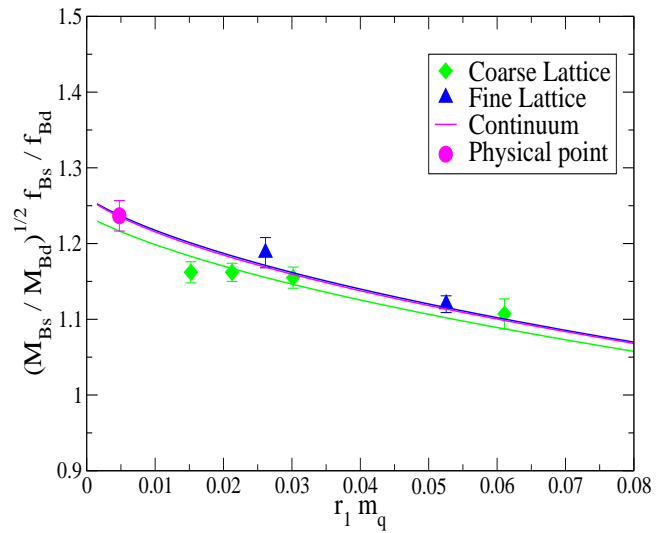}
\caption{
Chiral and continuum extrapolation of the ratio $\Phi_s/\Phi_d$. 
The different curves and the physical point have same meanings 
as in Fig.3. Here the red (continuum) curve is essentially on top 
of the blue (fine lattice) curve.
 }
\end{figure}

\begin{figure}
\includegraphics*[width=8.5cm,height=7.0cm]{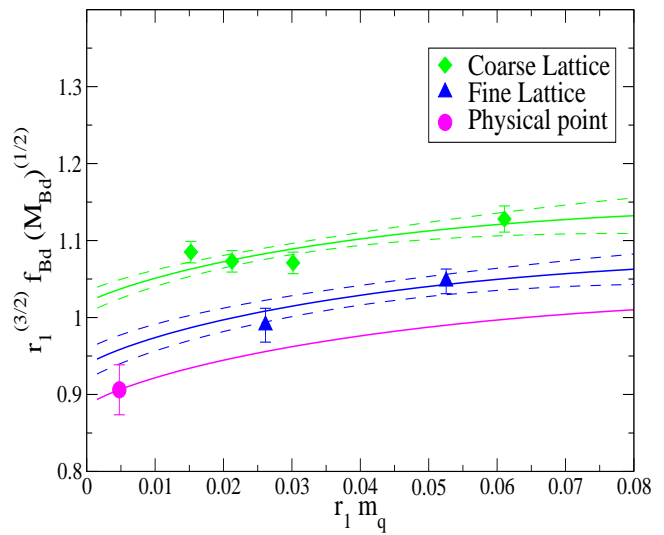}
\caption{
Same as Fig.4 for  $r_1^{3/2} \Phi_d = r_1^{3/2} f_{B_d} \sqrt{M_{B_d}}$. 
 }
\end{figure}

\begin{figure}[t]
\includegraphics*[width=8.5cm,height=7.0cm]{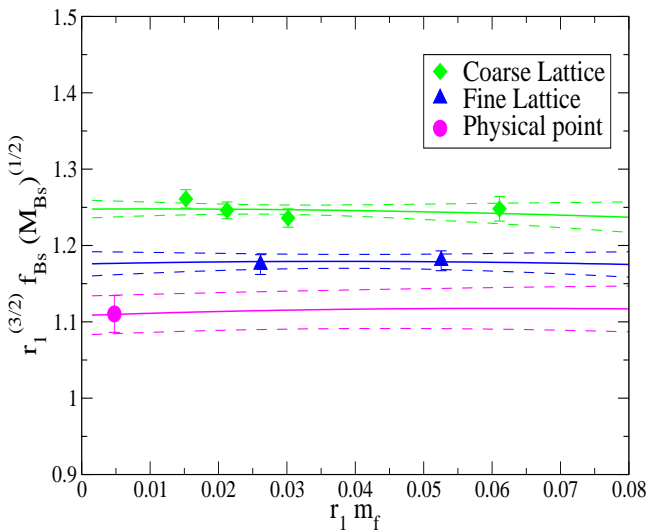}
\caption{
Same as Fig.5 for  $r_1^{3/2} \Phi_s = r_1^{3/2} f_{B_s} \sqrt{M_{B_s}}$. 
 }
\end{figure}

\begin{table}
\caption{ Errors in \% for 
$f_{B_s}$, $f_{B_d}$ and $f_{B_s} / f_{B_d}$.
}
\begin{center}
\begin{tabular}{|c|c|c|c|}
\hline
source of error & $\quad f_{B_s} \quad$ &
 $\quad f_{B_d} \quad $ & $\;\; f_{B_s} / f_{B_d} \;\;$ \\
\hline
\hline
stat. + chiral extrap.  & 2.2  &  3.5 &  1.6 \\
residual  $a^2$ extrap. &3.0  & 3.0 &  0.5\\
uncertainty &&&  \\
\hline
 $r_1^{3/2}$ uncertainty  & 2.3  &  2.3  &  --- \\
 $g_{B^*B\pi}$ uncertainty  & 1.0  &  1.0  &  0.3 \\
$m_s$ and $m_b$ tuning &  1.5  & 1.0  & 1.0  \\
operator matching  & 4.0 & 4.0& 0.7 \\
relativistic corr.  & 1.0 &  1.0 & 0.2\\
\hline
Total  & 6.3 & 6.7 & 2.1 \\
\hline
\end{tabular}
\end{center}
\end{table}

\noindent
Table V shows the error budget for $f_{B_s}$, $f_{B_d}$ and $f_{B_s} / f_{B_d}$, 
which is very similar to Table III for the mixing parameters. The meaning 
of the different sources of error is as explained in section V.  We have mentioned 
already that $\ord(\alpha_s \frac{\Lambda_{QCD}}{M})$ effects in the 
matching of the heavy-light current have been taken into account in our 
one-loop matching calculations \cite{pert1}. 
 Hence, these should not be included under 
 ``relativistic corrections'' in Table V.  However, there are still 
$\ord(\alpha_s \frac{\Lambda_{QCD}}{M})$ corrections to worry about 
in the NRQCD action.  These would come from radiative corrections to the 
coefficient $c_B$ (often also denoted $c_4$) 
of the $\frac{1}{2M}\,\sigmav\cdot\Bv$ term in the action.
 Although one-loop corrections to $c_B$ have not been calculated yet, 
one can nevertheless bound this coefficient nonperturbatively by 
calculating the hyperfine, the $B^*-B$, splitting and comparing with 
experiment.  Preliminary results discussed in \cite{upsilon} indicate that 
$c_B$ is close to one and the entire effect 
would  be at most  a 10\% correction to a $\frac{\Lambda_{QCD}}{M}$ 
contribution. 
 For the present calculations this means an uncertainty of order 1\%
in $f_{B_q}$ and a much smaller one for $\frac{f_{B_s}}{f_{B_d}}$.

\noindent
The final numbers for the decay constants including all errors 
 added in quadrature become,
\be
\label{fbrat}
\frac{f_{B_s}}{f_{B_d}} = 1.226(26),
\ee
\be
\label{fbd}
f_{B_d} = 
 190(13) \;\left (\frac{0.321}{r_1[fm]} \right)^{3/2} {\rm MeV},
\ee
and
\be
\label{fbs}
f_{B_s} = 
 231(15) \;\left (\frac{0.321}{r_1[fm]} \right)^{3/2} {\rm MeV}.
\ee
These results for $f_{B_q}$ are consistent with but about one $\sigma$ 
lower than the values $f_{B_d} = 216(22)$MeV 
and $f_{B_s} = 260(29)$MeV 
given in \cite{fbprl, fbsprl}.
 The main difference between the analysis carried out 
here and in \cite{fbprl} is that in the latter case chiral extrapolations were 
done based only on coarse lattice data and furthermore no attempt 
was made to extrapolate explicitly to the continuum limit.  
The new result for the ratio in (\ref{fbrat}) is similarly consistent with our 
previous $f_{B_s}/f_{B_d} = 1.20(3)(1)$ \cite{fbprl}. 

\vspace{.1in}
\noindent
The $B_q$ mixing simulations can also be used to determine the bag parameters 
$B_{B_q}$.  
From the separate final results for $f_{B_q} \sqrt{B_{B_q}}$ and 
for $f_{B_q}$ one 
 finds $B_{B_s}(\mu = M_b) = 0.86(6)$
 and $B_{B_d}(M_b) = 0.84(10)$.
We have also attempted to extract the bag parameters directly from 
simultaneous fits to 3-point and 2-point correlators for each ensemble 
separately.  The results are shown for $\hat{B}_{B_s}$ in Fig.10 and 
extrapolated to the physical point, where one finds
$\hat{B}_{B_s} = 1.33(5)$.  We add to this 3.8\% statistical + chiral extrapolation 
error additional 2.5\% systematic errors.
Many of the systematic errors listed in Tables 
III and V are either irrelevant or cancel to a large extent in $\hat{B}_{B_s}$.
For instance, one sees from Fig.10 that there is little evidence 
for discretization errors in the bag parameter,
 although $f_{B_s} \sqrt{M_{B_s} B_{B_s}}$ 
and $f_{B_s} \sqrt{M_{B_s}}$ individually do have noticeable lattice spacing 
dependence.  Our final result for the $B_s$ meson bag parameter is 
$\hat{B}_{B_s} = 1.33(6)$, or $B_{B_s}(M_b) = 0.86(4)$.  One sees that direct extraction 
of the bag parameter followed by a continuum/chiral extrapolation can 
reduce errors significantly over our first approach of extrapolating 
 $f_{B_s} \sqrt{M_{B_s} B_{B_s}}$ 
and $f_{B_s} \sqrt{M_{B_s}}$ first and then taking the ratio.
Although we were successful in applying the second method for $B_{B_s}$, unfortunately 
it has not been possible to get stable extractions of  $B_{B_d}$ for all of our 
six ensembles.  Our best estimate for the $B_d$ meson bag parameter is obtained 
by taking the two ratios, $f_{B_s}\sqrt{B_{B_s}}/f_{B_d}\sqrt{B_{B_d}}$ and 
$f_{B_s}/f_{B_d}$, from eqs.(21) and (31) and combining their ratio 
with our  most accurate $B_{B_s}(M_b)$.  This leads to $B_{B_s}/B_{B_d} 
= 1.05(7)$ and  $B_{B_d}(M_b) = 0.82(7)$.

\begin{figure}
\includegraphics*[width=8.5cm,height=7.0cm]{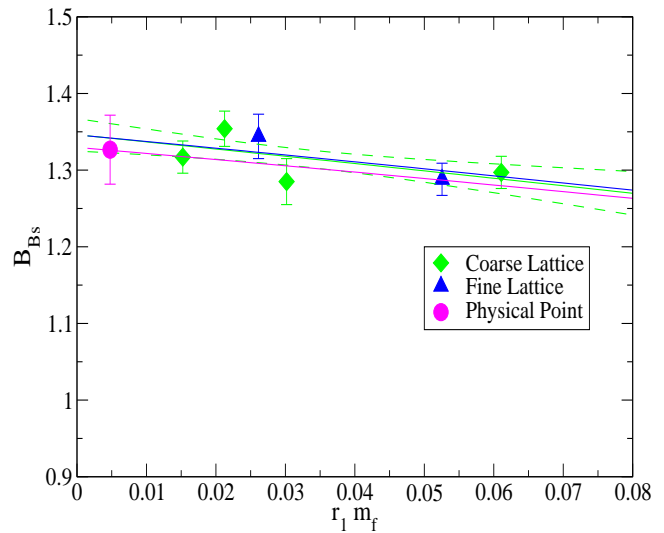}
\caption{
Chiral and continuum extrapolation of the bag parameter 
$\hat{B}_{B_s}$. In order to avoid clutter, we show an error band 
only for the fit to the coarse lattice data. 
 }
\end{figure}

\section{Summary}
We have completed the 
first $N_f = 2 + 1$ unquenched study of $B_s$ and $B_d$ mixing phenomena 
in Lattice QCD.  Our main results,
 namely values for $\xi = \frac{f_{B_s}\sqrt{B_{B_s}}}
{f_{B_d}\sqrt{B_{B_d}}}$, $f_{B_s} \sqrt{\hat{B}_{B_s}}$
 and $f_{B_d} \sqrt{\hat{B}_{B_d}}$, 
are given in eqns.(\ref{xires}), (\ref{fbsres}) and (\ref{fbdres}) 
respectively.  Combining the lattice result for $\xi$ with the experimentally 
measured mass differences $\Delta M_d = 0.507 \pm 0.005 \; ps^{-1}$ \cite{pdg} and 
$\Delta M_s = 17.77 \pm 0.10 \pm 0.07 \; ps^{-1}$ \cite{cdf} leads to,
\be
\frac{|V_{td}|}{|V_{ts}|} = 0.214(1)(5)
\ee
where the first error is experimental and the second from the 
lattice calculation presented here.  This is the first time this
ratio of CKM matrix elements 
has been determined while incorporating a fully self consistent $N_f=2+1$ 
calculation of $\xi$.

\vspace{.1in}
\noindent
In addition to giving  mixing parameter results, 
in this article  we have also 
updated values for decay constants $f_{B_d}$ and $f_{B_s}$ and their 
ratio in section VI.
$f_{B_s}$ appears in the Standard Model prediction for the
branching fraction for $B_s \rightarrow \mu^+ \mu^- $ \cite{bb}, 
a process sensitive to new physics. The most accurate
result for this branching fraction comes from taking a
ratio with $\Delta M_s$ \cite{buras2}, which gives a result
inversely proportional to $\hat{B}_{B_s}$. Updating the parameters
used in \cite{buras2} for 
$\tau(B_s)$ and $\overline{m_t}$ \cite{pdg}, and including our result
for
$\hat{B}_{B_s}$ of 1.33(6), gives
\be
Br(B_s \rightarrow \mu^+ \mu^-)
 = 3.19(19) \times 10^{-9},
\ee
improving on the previous accuracy available. The largest contribution to the error
on the branching fraction comes from the error on
$\hat{B}_{B_s}$ followed by the uncertainty in $\tau(B_s)$.

\vspace{.1in}
\noindent
The calculations presented here can be improved 
in several ways. 
Foremost among the improvements planned for the future is to 
carry out simulations at other, finer, lattice spacings. Having results 
at more than two lattice spacings will help considerably in reducing the 
``statistical  + chiral extrapolation'' and the ``residual $a^2$ 
extrapolation'' uncertainties in Tables III \& V.  They would also 
contribute to constraining the value of $g_{B^*B\pi}$ so that this 
source of error can then be ignored.  Hence, one can expect lattice results 
for $\xi$ (and also for $f_{B_s}/f_{B_d}$) with accuracy of $\sim 1$\% 
in the not too distant future.  Improvement for dimensionful quantities 
such as $f_{B_q} \sqrt{\hat{B}_q}$ will also require reducing the 
``$r_1^{3/2}$'' and the ``operator matching'' errors.  The HPQCD 
collaboration is currently engaged in projects aimed at fixing the 
physical value of $r_1$ \cite{hpqcd1} with higher precision than in the past. 
 We are also exploring nonperturbative 
methods for carrying out operator matching in heavy-light systems.  At least 
for heavy-light currents, methods recently applied to accurate determinations 
of heavy quark masses, which involve moments of current correlators 
and very high order continuum QCD perturbation theory 
\cite{hpqcd2}, look  promising 
for nonperturbative determinations of $Z$-factors.  More work will be 
necessary to see whether such methods can be generalized to 
four-fermion operators.  It is possible one can take advantage of the 
fact that a major contribution to matching of four-fermion 
operators comes from diagrams involving radiative corrections to just one 
of the bilinears within the four-fermion operator, in other words corrections 
that are identical to a heavy-light current radiative correction.  This has been 
noted already in the one-loop calculations of \cite{fourfmatch}.  With several of 
these improvements in place better than $\sim 5$\% accuracy should be achievable 
for $f_{B_q} \sqrt{\hat{B}_{B_q}}$.  

\noindent
Another worthwhile direction for future investigations would be to calculate 
hadronic matrix elements of further $\Delta B = 2$ four-fermion operators, beyond 
the two, $OL$ and $OS$, studied here.  As is well known, there are five such operators 
usually denoted $Qi$, with $i=1,2,3,4,5$ \cite{gabbiani,becirevic,fourfmatch}. 
 In this notation $OL \equiv Q1$ and 
$OS \equiv Q2$.  In this article we have  focused on $Q1$ and $Q2$ since only they 
are relevant for the mass difference $\Delta M_q$ in the Standard Model.  The 
operator $Q3$ would come in for calculations of the width difference 
$\Delta \Gamma_q$ \cite{lenz}.  Although we have already accumulated 
simulation data for $\langle Q3 \rangle$ we will postpone analysis for a 
future publication where we also plan to have results for $\langle Q4 \rangle$ 
and $\langle Q5 \rangle$. In \cite{fourfmatch} the necessary matching 
at one loop has already been completed for all five four-fermion operators. 
The two hadronic matrix elements $\langle Q4 \rangle$ and $\langle Q5 \rangle$ do not appear 
in the Standard Model but are of interest in several Supersymmetric Models.
To date only quenched lattice results exist for all five four-fermion operators 
\cite{becirevic}. It will be important for Beyond the Standard Model studies to 
generalize those results to unquenched calculations.

\vspace{.1in}
{\bf Acknowledgments}: \\
This research was supported by the DOE and NSF (USA), the STFC,
 Royal Society and Leverhulme Trust (UK), and by the Junta de 
Andaluc\'{\i}a  (Spain). 
 Numerical work 
was carried out on facilities of the USQCD Collaboration funded by the Office 
of Science of the U.S. DOE. We are grateful to the MILC collaboration for making their
gauge configurations available and for updates on $r_1/a$.  We thank Claude Bernard, 
Jack Laiho and Ruth Van de Water for providing 
 their results on Staggered ChPT  for 
four-fermion operators prior to publication.


\begin{thebibliography}{99}
\bibitem{bayes2}
See reference \cite{bayes} for a definition of $\chi^2_{aug}$.  
Note also that with Bayesian fits ``dof'' equals the number of 
data points (and not number of data points minus the number of fit parameters) 
as explained after eq.(8) of this same reference.
\end{thebibliography}


\end{document}